\documentclass{llncs}

\usepackage{t1enc} 
\usepackage{amsmath}
\usepackage{latexsym}
\usepackage{theorem}
\usepackage{amssymb}
\usepackage{url}

\newcommand{\comment}[1]{}

\newcommand{\hsp}[1][3mm]{\hspace*{#1}}
\newcommand{\vsp}[1][3mm]{\vspace*{#1}}
\newcommand{\moins}{\setminus}
\newcommand{\vide}{\emptyset}
\newcommand{\ie}{{\em i.e.} }
\newcommand{\eg}{{\em e.g.} }

\newcommand{\dom}{\mr{dom}}

\newcommand{\FV}{\mr{FV}}

\renewcommand{\a}{\rightarrow}
\newcommand{\A}{\Rightarrow}

\renewcommand{\AA}{\Leftrightarrow}

\newcommand{\ad}{\downarrow}

\newcommand{\au}{\uparrow}
\renewcommand{\to}{\mapsto}

\newcommand{\I}[1]{[\![#1]\!]}
\newcommand{\J}[1]{(\!|#1|\!)}

\newcommand{\ex}{\exists}
\newcommand{\all}{\forall}
\newcommand{\ou}{\vee}

\newcommand{\et}{\wedge}

\renewcommand{\th}{\vdash}

\newcommand{\sle}{\subseteq}

\newcommand{\sgt}{\supset}

\newcommand{\lex}{_\mr{lex}}

\renewcommand{\o}[1]{{\overline{#1}}}

\renewcommand{\b}{\beta}
\newcommand{\g}{\gamma}
\newcommand{\G}{\Gamma}
\renewcommand{\d}{\delta}
\newcommand{\D}{\Delta}

\newcommand{\vep}{\varepsilon}

\renewcommand{\t}{\theta}

\newcommand{\kap}{\kappa}
\renewcommand{\l}{\lambda}
\renewcommand{\L}{\Lambda}

\newcommand{\s}{\sigma}
\renewcommand{\S}{\Sigma}

\newcommand{\vphi}{\varphi}
\newcommand{\w}{\omega}

\newcommand{\mc}{\mathcal}

\newcommand{\mr}{\mathrm}
\newcommand{\mb}{\mathbb}

\newcommand{\mf}{\mathfrak}
\newcommand{\ms}{\mathsf}

\newcommand{\bB}{\mb{B}}
\newcommand{\bC}{\mb{C}}

\newcommand{\bE}{\mb{E}}

\newcommand{\bN}{\mb{N}}

\newcommand{\bS}{\mb{S}}
\newcommand{\bT}{\mb{T}}

\newcommand{\cA}{\mc{A}}
\newcommand{\cB}{\mc{B}}
\newcommand{\cC}{\mc{C}}
\newcommand{\cD}{\mc{D}}
\newcommand{\cE}{\mc{E}}
\newcommand{\cF}{\mc{F}}

\newcommand{\cN}{\mc{N}}

\newcommand{\cP}{\mc{P}}

\newcommand{\cR}{\mc{R}}
\newcommand{\cS}{\mc{S}}
\newcommand{\cT}{\mc{T}}

\newcommand{\cX}{\mc{X}}

\newcommand{\ka}{\mf{a}} 
\newcommand{\kb}{\mf{b}}

\newcommand{\kd}{\mf{d}}

\newcommand{\kf}{\mf{f}}

\newcommand{\kt}{\mf{t}}

\newcommand{\fb}{\ms{b}}
\newcommand{\fc}{\ms{c}}

\newcommand{\ff}{\ms{f}}
\newcommand{\fg}{\ms{g}}

\newcommand{\fs}{\ms{s}}

\newcommand{\fB}{\ms{B}}
\newcommand{\fC}{\ms{C}}

\newcommand{\va}{{\vec{a}}}
\newcommand{\vb}{{\vec{b}}}

\newcommand{\vl}{{\vec{l}}}

\newcommand{\vp}{{\vec{p}}}

\newcommand{\vt}{{\vec{t}}}
\newcommand{\vu}{{\vec{u}}}

\newcommand{\vx}{{\vec{x}}}

\newcommand{\vP}{{\vec{P}}}

\newcommand{\vT}{{\vec{T}}}
\newcommand{\vU}{{\vec{U}}}

\newenvironment{rul}[1][~~\a~~]
  {$\begin{array}{r@{#1}ll}}
  {\end{array}$}
\newenvironment{rulc}[1][~~\a~~]
  {\begin{center}\begin{rul}[#1]}
  {\end{rul}\end{center}}

\newenvironment{trans}[1][~~\a~~]
  {\begin{center}$\begin{array}{rr@{#1}ll}}%
  {\end{array}$\end{center}}

{\theorembodyfont{\rmfamily} 

}

\newenvironment{lstgeneric}[2]
  {\begin{list}{#1}{\topsep=.5mm\itemsep=.5mm\parsep=0mm%
    \itemindent=-3ex\labelsep=1ex\labelwidth=0ex #2}}
  {\end{list}}

\newenvironment{lst}[1]
  {\begin{lstgeneric}{#1}{\itemindent=-1ex}}
  {\end{lstgeneric}}

\newenvironment{enumi}[1]
  {\begin{lstgeneric}{}{\usecounter{enumi}\leftmargin=7mm%
    }}
  {\end{lstgeneric}}

\newcommand{\SN}{\mr{SN}}
\newcommand{\SAT}{\mr{SAT}}

\newcommand{\liste}{{\sf L}}
\newcommand{\nat}{{\sf N}}
\newcommand{\tree}{{\sf T}}
\newcommand{\bool}{{\sf bool}}

\newcommand{\open}{{\sf let\,}}
\newcommand{\dans}{{\sf\,in\,}}
\newcommand{\si}{{\sf if\,}}
\newcommand{\alors}{{\sf\,then\,}}
\newcommand{\sinon}{{\sf\,else\,}}
\newcommand{\fst}{{\sf fst\,}}
\newcommand{\snd}{{\sf snd\,}}

\newcommand{\plus}{{\sf plus\,}}
\newcommand{\filter}{{\sf filter\,}}

\newcommand{\pivot}{{\sf pivot\,}}
\newcommand{\app}{{\sf append\,}}
\newcommand{\nil}{{\sf nil\,}}
\newcommand{\cons}{{\sf cons\,}}
\newcommand{\node}{{\sf node\,}}
\newcommand{\leaf}{{\sf leaf\,}}

\newcommand{\minus}{{\sf minus\,}}
\renewcommand{\div}{{\sf div\,}}
\newcommand{\zero}{{\sf 0\,}}
\newcommand{\true}{{\sf true}}
\newcommand{\false}{{\sf false}}
\renewcommand{\inf}{{\sf le\,}}

\newcommand{\ar}{\a_\cR}
\newcommand{\ab}{\a_\b}
\newcommand{\abh}{\a_{\b h}}
\newcommand{\abwh}{\a_{\b wh}}

\newcommand{\red}[1]{{\a\!\!(#1)}}

\newcommand{\nf}[1]{{#1\!\ad}}

\newcommand{\valpha}{{\vec\alpha}}
\newcommand{\vbeta}{{\vec\b}}

\newcommand{\vdelta}{{\vec\d}}

\newcommand{\vka}{{\vec\ka}}

\newcommand{\dtrue}{{\kt}}
\newcommand{\dfalse}{{\kf}}
\newcommand{\atrue}{\mr{t}}
\newcommand{\afalse}{\mr{f}}

\begin{document}


\title{Combining typing and size constraints for checking the termination of higher-order conditional rewrite systems}

\author{Fr\'ed\'eric Blanqui (INRIA) \and Colin Riba (INPL)}

\institute{LORIA\thanks{UMR 7503 CNRS-INPL-INRIA-Nancy2-UHP},
Campus Scientifique, BP 239\\
54506 Vandoeuvre-l\`es-Nancy Cedex, France}

\maketitle

\begin{abstract}
In a previous work, the first author extended to higher-order
rewriting and dependent types the use of size annotations in types, a
termination proof technique called type or size based termination and
initially developed for ML-like programs. Here, we go one step further
by considering conditional rewriting and explicit quantifications and
constraints on size annotations. This allows to describe more
precisely how the size of the output of a function depends on the size
of its inputs. Hence, we can check the termination of more
functions. We first give a general type-checking algorithm based on
constraint solving. Then, we give a termination criterion with
constraints in Presburger arithmetic. To our knowledge, this is the
first termination criterion for higher-order conditional rewriting
taking into account the conditions in termination.
\end{abstract}


\section{Introduction}

We are interested in automatically checking the termination of the
combination of $\b$-reduction and higher-order conditional rewrite
rules. There are two important approaches to higher-order rewriting:
rewriting on $\b\o\eta$-normal forms \cite{mayr98tcs}, and the
combination of $\b$-reduction and term rewriting \cite{klop93tcs}. The
relation between both has been studied in \cite{oostrom93hoa}. The
second approach is more atomic since a rewrite step in the first
approach can be directly encoded by a rewrite step together with
$\b$-steps in the second approach. In this paper, we consider the
second approach, restricted to first-order pattern-matching (we do not
allow abstractions in rule left-hand side). Following
\cite{blanqui00rta}, our results could perhaps be extended to
higher-order pattern-matching.

The combination of $\b$-reduction and rewriting is naturally used in
proof assistants implementing the proposition-as-type and
proof-as-object paradigm. In these systems, two propositions
equivalent modulo $\b$-reduction and rewriting are identified (\eg
$P(2+2)$ and $P(4)$). This is essential for enabling users to
formalize large proofs with many computations, as recently shown by
Gonthier and Werner's proof of the Four Color Theorem in the Coq proof
assistant. However, for the system to be able to check the correctness
of user proofs, it must at least be able to check the equivalence of
two terms. Hence, the necessity to have termination criteria for the
combination of $\b$-reduction and rewriting.

In Coq, rewriting is restricted to the reductions associated to
inductive types like in functional programming languages with
pattern-matching. Such reductions correspond to constructor-based
rewriting. This is the kind of rewrite systems we are going to
consider in this paper. A more general form of rewriting is studied in
\cite{blanqui05mscs,blanqui03rta} (matching on defined symbols and
matching modulo).

Currently, Coq accepts only functions in the definition of which
recursive calls are made on arguments that are structurally
smaller. For first-order functions, this corresponds to restrict
rewrite systems to simply terminating ones, that is, to the ones that
can be proved terminating by an ordering containing the subterm
relation. However, many interesting systems are not simply
terminating. Consider for instance the following definition of
division on natural numbers:

\begin{rulc}
\minus \zero x & \zero\\
\minus x\,\zero & x\\
\minus (\fs\,x)\,(\fs\,y) & \minus x\,y\\
\div \zero\,y & \zero\\
\div (\fs\,x)\,y & \fs\,(\div (\minus x\,y)\,y)\\
\end{rulc}

Considering that $\minus$ is applied to strongly normalizing arguments
and that the {\em size} of a term is the height of its normal form,
one can easily prove, by induction on the size of $t$, that the size
of $v=(\minus t\,u)$ is less than or equal to the size of $t$, hence
that this definition of $\minus$ terminates:

\begin{lst}{--}
\item If $v$ matches the first rule, then $t=\ms{0}$ and the normal
form of $v$, which is $\ms{0}$, has the same size as $t$.
\item If $v$ matches the second rule, then $v$ has the same normal
form as $t$.
\item If $v$ matches the third rule, then $t=\fs t'$, $u=\fs u'$
and, by induction hypothesis, the normal form of $v$ has a size
smaller than $t'$, hence smaller than $t$.
\end{lst}

The idea of size or type based termination, initiated in
\cite{hughes96popl} and developed by various authors for ML-like
definitions
\cite{chin01hosc,xi02hosc,abel03tlca,abel04ita,barthe04mscs,barthe05tlca}
and rewriting and dependent types \cite{blanqui04rta,blanqui05csl},
consists in extending the underlying type system by replacing a base
type $\fB$ by an infinite family of base types
$(\fB^\ka)_{\ka\in\bN}$, a term of type $\fB^\ka$ being by
construction of size smaller than or equal to $\ka$ (except in
\cite{xi02hosc}, see later). Then, for ensuring termination, one can
restrict in function definitions recursive calls to arguments whose
size, by typing, is smaller.

For instance, in all these systems, one can easily (type-)check that
$\minus$ has for type something similar to
$\all\alpha\b\nat^\alpha\A\nat^\b\A\nat^\alpha$. Hence, assuming that
$x:\nat^\alpha$ and $y:\nat^\b$, one can easily (type-)check that
$\minus x\,y:\nat^\alpha$ while $\fs x:\nat^{\alpha+1}$. Thus, the
recursive call to $\div$ in the last rule can be allowed.

Note that higher-order inductive types, \ie types having constructors
with recursive arguments of higher-order type, require families
indexed by ordinals. In the present paper, we restrict our attention
to first-order inductive types since higher-order inductive types have
already been studied in previous works. Note also that interpreting
$\fB^\ka$ by the set of terms of size smaller than or equal to $\ka$
requires subtyping since $t:\fB^\kb$ whenever $t:\fB^\ka$ and
$\ka\le\kb$.

However, without explicit existential quantifications and constraints
over size annotations, one cannot (type-)check that the following
function has type
$\nat\A\all\alpha\liste^\alpha\A\ex\b\g(\alpha=\b+\g)\liste^\b\times\liste^\g$:

\begin{center}
$\begin{array}{rcll}
\pivot x\,\nil &~~\a~~& (\nil,\nil)\\
\pivot x\,(\cons y\,l) &~~\a~~&
\open z=\pivot x\,l\dans\\
&& \si (\inf y\,x)\alors (\cons y\,(\fst z),\snd z)\\
&& \sinon (\fst z,\cons y\,(\snd z))\\
\end{array}$
\end{center}

Such a type is necessary for proving that some sorting functions
are size preserving, \ie have type
$\all\alpha\liste^\alpha\A\liste^\alpha$. To the best of our
knowledge, only Xi considers such explicit quantifications and
constraints \cite{xi02hosc}. In this work, $\fB^\ka$ is interpreted as
the set of terms of size $\ka$. Note that, with this interpretation,
the type of terms of size smaller than $\ka$ can be represented by
$\ex\alpha(\alpha\le\ka)\fB^\alpha$. However, we cannot apply Xi's
results on the problem we are interested in for the following reasons:

\begin{lst}{--}
\item Xi considers ML-like function definitions based on {\tt letrec}/{\tt match}
constructions while we are interested in definitions based on rewrite rules.

\item Xi is interested in the termination of closed terms with call-by-value
evaluation strategy while we are interested in the strong
normalization of open terms.

\item Xi has a two-level approach. He considers an intermediate system where
not only types but also terms are annotated by size informations, and
proves that terms typable in this system are terminating. Then, for
proving the termination of an unannotated term, he must infer the
necessary size annotations, which may not be possible. This
elaboration process is described in \cite{xi98thesis}.
\end{lst}

In the present paper, we extend the simply typed part of
\cite{blanqui04rta} with conditional rewriting and explicit
quantifications and constraints over size annotations, without using
an intermediate system. As Xi and in contrast with
\cite{blanqui04rta}, we do not consider higher-order inductive types
and interpret $\fB^\ka$ as the set of terms of size $\ka$. The
integration of both works should not create too much
difficulties. Hence, we get a powerful termination criterion for the
combination of $\b$-reduction and higher-order conditional rewriting,
based on type-checking and constraint solving. To our knowledge, this
is the first termination criterion for higher-order conditional
rewriting taking into account the conditions in termination.

In Section \ref{sec-def}, we define a system with constrained
types. In Section \ref{sec-type-check}, we give a general
type-checking algorithm based on constraint solving. In Section
\ref{sec-sn}, we present a general termination proof technique based
on Tait's method for proving the termination of $\b$-reduction. In
Section \ref{sec-crit}, we give a termination criterion based on
type-checking with constraints in Presburger arithmetic.


\section{A system with constrained types}
\label{sec-def}

{\bf Terms.} The set $\cT$ of {\em terms} is inductively defined as
follows:

\begin{center}
$t\in\cT ::= x ~|~ \fc ~|~ \ff ~|~ \l xt ~|~ tt
~|~ (t,t) ~|~ \fst t ~|~ \snd t ~|~ \open x=t\dans t
~|~ \si t\alors t\sinon t$
\end{center}

\noindent
where $x\in\cX$ is a term variable, $\fc\in\cC$ is a {\em constructor}
symbol and $\ff\in\cF$ is a {\em function symbol}. We assume that
$\cC$ contains $\true$ and $\false$. As usual, terms are considered up
to renaming of bound variables. By $\vt$, we denote a sequence of
terms $t_1,\ldots,t_n$ of length $|\vt|=n\ge 0$. Term substitutions
are denoted by $\s,\t,\ldots$ or their explicit mappings
$(_\vx^\vt)$. By $\s+\t$, we denote the substitution equal to $\t$ on
$\dom(\t)$ and to $\s$ on $\dom(\s)\moins\dom(\t)$. The set $\cP$ of
(constructor) {\em patterns} is inductively defined by $p\in\cP ::= x
~|~ \fc\,\vp$.

{\bf Size annotations.} Let $\cS=\{nat,bool\}$ be the set of {\em size
sorts}. We assume given a $\cS$-sorted first-order term algebra $\cA$
for {\em size expressions} $a,b,\ldots$ whose variables are denoted by
$\alpha,\b,\ldots$ We assume that $\cA$ at least contains the symbols
$0:nat$, $1:nat$, $+:nat\times nat\A nat$, $max:nat\times nat\A nat$,
$\atrue:bool$ and $\afalse:bool$. For each sort $s$, we assume given a
well-founded interpretation domain $(\cD_s,>_{\cD_s})$. For $bool$, we
take $\cD_{bool}=\{\dtrue,\dfalse\}$. In the following, let
$\true^*=\atrue$ and $\false^*=\afalse$; $\atrue^*=\dtrue$ and
$\afalse^*=\dfalse$; $\dtrue^*=\true$ and $\dfalse^*=\false$. Elements
of $\cD_s$ are denoted by $\ka,\kb,\ldots$ Valuations are denoted by
$\mu,\nu,\ldots$ Size substitutions are denoted by $\vphi,\psi,\ldots$

{\bf Constraints.} Let a {\em constraint} be a first-order formula
over $\cA$, $\bC$ be a class of constraints containing $\top$ and
$\FV(C)$ be the variables free in $C$. We denote by $\mu\models C$ the
fact that a valuation $\mu$ satisfies $C$; by $\th C$ the fact that,
for all valuation $\mu$ such that $\FV(C)\sle\dom(\mu)$, $\mu\models
C$, and by $C\equiv D$ the fact that $\th C\AA D$. We consider
constraints up to the logical equivalence $\equiv$.

{\bf Types.} We assume given a set $\cB$ of type names containing
$\bool$. Let $\kap_\bool=bool$ and, for all $\fB\neq\bool$,
$\kap_\fB=nat$ (except $\bool$ that is annotated by booleans, types
are annotated by natural numbers). Types are defined as follows:

\begin{center}
\begin{tabular}{rr@{~::=~}l}
types & $T\in\bT$
& $\fB^a ~|~ T\A T ~|~ T\times T ~|~ \all\valpha PT ~|~ \ex\valpha PT$\\
simple types & $S\in\bS$ & $\ex\alpha\fB^\alpha ~|~ S\A S ~|~ S\times S$\\
basic types & $B\in\bB$ & $\fB^a ~|~ B\times B$\\
$\ex$-basic types & $E\in\bE$
& $B ~|~ \ex\valpha P E\quad\mbox{with } \th\ex\valpha P$\\
\end{tabular}
\end{center}

\noindent
where $\fB\in\cB$ is a type name, $a\in\cA$ is a size expression of
sort $\kap_\fB$ and $P\in\bC$ is a constraint. In the following, we
use the following abbreviations: $\all\valpha T=\all\valpha\top T$ and
$\fB=\ex\alpha\fB^\alpha$. There is a natural transformation from
$\bT$ to $\bS$: let $\o{\fB^a}=\ex\alpha\fB^\alpha$, $\o{\ex\alpha
PT}=\o{\all\alpha PT}=\o{T}$, $\o{T\A U}=\o{T}\A\o{U}$ and $\o{T\times
U}=\o{T}\times\o{U}$.

{\bf Subtyping.} We define a constraint-based subtyping relation. Let
$C\th T\le U$ iff $\th C\sgt\J{T\le U}$ where $\J{T\le U}$ is
inductively defined as follows:

\begin{lst}{--}
\item $\J{B^a\le B^b}= (a=b)$
\item $\J{T\A U\le T'\A U'}= \J{T'\le T}\et \J{U\le U'}$
\item $\J{T\times U\le T'\times U'}= \J{T\le T'}\et \J{U\le U'}$
\item $\J{T\le \ex\valpha PU}= \ex\valpha(P\et\J{T\le U})$
($\valpha\notin T$, $T\neq\ex\b QV$)
\item $\J{\ex\valpha PU\le T}= \all\valpha(P\sgt\J{U\le T})$ ($\valpha\notin T$)
\item $\J{T\le \all\valpha PU}= \all\valpha(P\sgt\J{T\le U})$ ($\valpha\notin T$)
\item $\J{\all\valpha PU\le T}= \ex\valpha(P\et\J{U\le T})$
($\valpha\notin T$, $T\neq\all\b QV$)
\end{lst}

{\bf Typing.} An {\em environment} is a finite mapping $\G$ from $\cX$
to $\bT$. Let $\G,x:T$ be the environment $\D$ such that $x\D=T$ and
$y\D=y\G$ if $y\neq x$. Two environments $\G_1$ and $\G_2$ are {\em
compatible} if, for all $x$, $x\G_1=x\G_2$.

A type assignment is a function $\tau:\cC\cup\cF\a\bT$ such that
$\tau_\true=\bool^\atrue$, $\tau_\false=\bool^\afalse$ and, for all
$\fs\in\cC\cup\cF$, $\tau_\fs$ is closed. To every type assignment
$\tau$, we associate a typing relation $\th_\tau$ defined in Figure
\ref{fig-typ}. Note that, in contrast with
\cite{xi02hosc}, the typing of $u$ and $v$ in (if) does not depend on
$t$. This is because we consider strong normalization instead of weak
normalization. This does not reduce the expressive power of the system
since we consider conditional rewriting.

A term $t$ is typable wrt $\tau$ if there are $C,\G,T$ such that $\th
C$ and $C;\G\th_\tau t:T$. Let $\L(\tau)$ be the set of terms typable
wrt $\tau$. A term $t$ is {\em simply typable} if there are $\G,T$
simple such that $\top;\G\th_{\o\tau} t:T$ without ($\ex$intro),
($\all$intro), ($\ex$elim), ($\all$elim), (sub). Let $\o\L(\o\tau)$ be
the set of terms simply typable wrt $\o\tau$.


\begin{figure}[ht]
\centering\caption{Typing rules\label{fig-typ}}\vsp
\fbox{\begin{minipage}{12cm}\centering\vsp
(var)\quad
$\cfrac{x\in\dom(\G)}
{C;\G\th_\tau x:x\G}$\hsp[5mm]
(symb)\quad
$\cfrac{\fs\in\cC\cup\cF}{C;\G\th_\tau \fs:\tau_\fs}$\\[2mm]

(abs)\quad
$\cfrac{C;\G,x:T\th_\tau u:U\quad x\notin\G}
{C;\G\th_\tau \l xu:T\A U}$\hsp[5mm]
(app)\quad
$\cfrac{C;\G\th_\tau t:U\A V\quad C;\G\th_\tau u:U}
{C;\G\th_\tau tu:V}$\\[2mm]

(pair)\quad
$\cfrac{C;\G\th_\tau u:U\quad C;\G\th_\tau v:V}
{C;\G\th_\tau (u,v):U\times V}$\\[2mm]

(fst)\quad
$\cfrac{C;\G\th_\tau t:U\times V}
{C;\G\th_\tau\fst t:U}$\hsp[5mm]
(snd)\quad
$\cfrac{C;\G\th_\tau t:U\times V}
{C;\G\th_\tau\snd t:V}$\\[2mm]

(if)\quad
$\cfrac{C;\G\th_\tau t:\bool\quad C;\G\th_\tau u:T
\quad C;\G\th_\tau v:T\quad T~\ex\mbox{-basic}}
{C;\G\th_\tau \si t\alors u\sinon v:T}$\\[2mm]

(let)\quad
$\cfrac{C;\G\th_\tau t:T\quad C;\G,x:T\th_\tau u:U\quad x\notin\G}
{C;\G\th_\tau \open x=t\dans u:U}$\\[2mm]

($\all$intro)\quad
$\cfrac{C\et P;\G\th_\tau t:T\quad \th C\sgt\ex\valpha P
\quad \valpha\notin C,\G}
{C;\G\th_\tau t:\all\valpha PT}$\\[2mm]

($\all$elim)\quad
$\cfrac{C;\G\th_\tau t:\all\valpha PT\quad \th C\sgt P_\valpha^\va}
{C;\G\th_\tau t:T_\valpha^\va}$\\[2mm]

($\ex$intro)\quad
$\cfrac{C;\G\th_\tau t:T_\valpha^\va\quad \th C\sgt P_\valpha^\va}
{C;\G\th_\tau t:\ex\valpha PT}$\\[2mm]

($\ex$elim)\quad
$\cfrac{C;\G\th_\tau t:\ex\valpha PT\quad C\et P;\G,x:T\th_\tau u:U
\quad \th C\sgt\ex\valpha P\quad \valpha,x\notin C,\G,U}
{C;\G\th_\tau \open x=t\dans u:U}$\\[2mm]

(sub)\quad
$\cfrac{C;\G\th_\tau t:T\quad C\th T\le T'}
{C;\G\th_\tau t:T'}$
\vsp\end{minipage}}
\end{figure}


\begin{example}
\label{ex-append}
Consider the symbols
$\app:\all\b\g\liste^\b\A\liste^\g\A\liste^{\b+\g}$ and
$\pivot:\nat\A\all\alpha\liste^\alpha\A\ex\b\g(\alpha=\b+\g)\liste^\b\times\liste^\g$.
Let $\G={x:\nat}$, ${l:\liste^\alpha}$, $u=(\open z=t\dans v)$,
$t=\pivot x\,l$ and $v=\app(\fst z)(\snd z)$. Then, $\top;\G\th
t:\ex\b\g(\alpha=\b+\g)\liste^\b\times\liste^\g$ and ${\alpha=\b+\g}$;
$\G,{z:\liste^\b\times\liste^\g}\th v:\liste^\alpha$. Thus, by
($\ex$elim), $\G\th u:\liste^\alpha$.
\end{example}


\comment{
\begin{lemma}
\begin{lst}{--}
\item If $C;\G\th t:T$ and $\th D\sgt C$, then $D;\G\th t:T$.
\item If $C;\G\th t:T$ and $\vphi$ is a size substitution, then
$C\vphi;\G\vphi\th t:T\vphi$ with a derivation of the same height.
\item If $C;\G\th t:T$ and $C;\D\th x\s:x\G$ whenever $x\in\dom(\G)$,
then $C;\D\th t\s:T$.
\end{lst}
\end{lemma}
}


{\bf Rewriting.} Let $\ab$ be the smallest relation stable by context
containing the {\em head-$\b$-reduction relation} $\abh$ defined as
follows:

\begin{center}
$\begin{array}{r@{~\abh~}l}
(\l xu)t & u_x^t\\
\open x=t\dans u  & u_x^t\\
\end{array}
\quad
\begin{array}{r@{~\abh~}l}
\fst(u,v) & u\\
\snd(u,v) & v\\
\end{array}
\quad
\begin{array}{r@{~\abh~}l}
\si\true\alors u\sinon v & u\\
\si\false\alors u\sinon v & v\\
\end{array}$
\end{center}

A {\em conditional rewrite rule} is an expression of the form
$\vt=\vec\fc\sgt l\a r$ such that $l$ is of the form $\ff\vl$, $\vl$
are patterns, $\vec\fc\in\{\true,\false\}$ and
$\FV(r,\vt)\sle\FV(l)$. A rule $\vt=\vec\fc\sgt l\a r$ {\em defines}
$\ff\in\cF$ if $l$ is of the form $\ff\vl$. In the following, we
assume given a set $\cR$ of rules. The associated rewrite relation is
the smallest relation $\ar$ stable by context and substitution such
that, for all $\vt=\vec\fc\sgt l\a r\in\cR$, $l\s\ar r\s$ whenever
$\vt\s\a^*\vec\fc$, where $\a^*$ is the reflexive and transitive
closure of $\a=\ab\cup\ar$.

Our goal is to prove the strong normalization of $\a=\ab\cup\ar$ on
the set of simply typable terms $\o\L(\o\tau)$.\\

{\bf Assumption:} We assume that $\a$ is locally confluent.\\

Hence, any strongly normalizing term $t$ has a unique normal form
$\nf{t}$. Note that $\a$ is locally confluent whenever $\ar$ so is.
See \cite{blanqui06fossacs} for general conditions on the confluence
of $\b$-reduction and higher-order conditional rewriting.

It should be noted that ($\ex$elim) makes subject reduction fail. For
instance, with $\G=x:\ex\alpha\nat^\alpha,
y:\all\alpha\nat^\alpha\A\ex\b\nat^\b$, we have $\top;\G\th \open
z=x\dans yz:\ex\b\nat^\b$ while $yx$ is not typable in $\top;\G$. It
could be fixed by replacing in ($\ex$elim) $\open x=t\dans u$ by
$u_x^t$. It does not matter since our termination proof technique does
not need subject reduction. Note however that subject reduction holds
on simply typed terms.

An example of higher-order conditional rule is given by the following
definition of
$\filter:(\nat\A\nat)\A\all\alpha\liste^\alpha\A\ex\b(\b\le\alpha)\liste^\b$:

\begin{trans}
& \filter f\,\nil & \nil\\
f\,x=\true\sgt & \filter f(\cons x\,l) & \cons (f\,x)\, (\filter f\,l)\\
f\,x=\false\sgt & \filter f(\cons x\,l) & \filter f\,l\\
\end{trans}


\section{Type-checking algorithm}
\label{sec-type-check}

Type-checking is the following problem: given $\tau$, $C$, $\G$, $t$
and $T$, do we have $C$ satisfiable and $C;\G\th_\tau t:T$ ?

Because of the rules ($\ex$elim) and (conv), type-checking does not
seem to be decidable. Similarly, in \cite{xi02hosc}, the elaboration
process is not complete. It is however possible to give an algorithm
that either succeed or fails, a failure meaning that we don't know.
To this end, we inductively define in Figure 2 two relations in the
style of bi-directional type inference
\cite{davies00icfp,abel04ita}. In the type inference relation $C;\G\th
t\au T$, $C$ and $T$ are produced according to $\G$ and $t$. In the
type checking relation $C;\G\th t\ad T$, $C$ is produced according to
$\G$, $t$ and $T$. An actual algorithm is a strategy for applying the
rules defining these relations.

Let $\o\bC$ be the closure of $\bC$ by conjunction, implication,
existential and universal quantification. If one starts with
$C\in\bC$, then the constraints generated by such an algorithm are in
$\bC$ too. Hence, if $\bC$ only contains linear inequalities, then
$\o\bC$ are formulas of Presburger arithmetic which is known to be
decidable \cite{presburger29} and whose complexity is doubly
exponential in the size of the formula \cite{fischer74sam}. This high
complexity is not so important in our case since the terms we intend
to consider are small (rule right-hand sides). It would be however
interesting to study in more details the complexity of type-checking
wrt $\bC$.

For proving the correctness of the rule ($\ad$$\ex$intro), we need to
assume that the size expression language $\cA$ is complete wrt the
interpretation domains $\cD_s$, that is, to every $\ka\in\cD_s$
corresponds a closed term $a\in\cA$ whose denotation in $\cD_s$ is
$\ka$. Note that this is indeed the case when $\cD_s=\bN$ and $\cA$
contains $0$, $1$ and $+$.

See Example \ref{ex-pivot} at the end of the paper for an example of
derivation.


\begin{figure}
\begin{center}
\caption{Rules for deciding type-checking\label{fig-infer}}\vsp
\fbox{\begin{minipage}{12cm}\centering\vsp
(type-check)\quad
$\cfrac{D;\G\th t\ad T\quad \th C\sgt D\quad C\mbox{ satisfiable}}
{C;\G\th^? t:T}$\\[5mm]

($\au$var)\quad
$\cfrac{x\in\dom(\G)}
{\top;\G\th x\au x\G}$\hsp[5mm]
($\au$symb)\quad
$\top;\G\th \fs\au\tau_\fs$\\[2mm]

($\au$app)\quad
$\cfrac{C;\G\th t\au U\A V\quad D;\G\th u\ad U}
{C\et D;\G\th tu\au V}$\\[2mm]

($\au$pair)\quad
$\cfrac{C;\G\th u\au U\quad D;\G\th v\au V}
{C\et D;\G\th (u,v)\au U\times V}$\\[2mm]

($\au$fst)\quad
$\cfrac{C;\G\th t\au U\times V}
{C;\G\th \fst t\au U}$\hsp[5mm]
($\au$snd)\quad
$\cfrac{C;\G\th t\au U\times V}
{C;\G\th \snd t\au V}$\\[2mm]

($\au$let)\quad
$\cfrac{C;\G\th t\au T\quad D;\G,x:T\th u\au U}
{C\et D;\G\th \open x=t\dans u\au U}$\\[2mm]

($\au$$\all$elim)\quad
$\cfrac{C;\G\th t\au \all\valpha PT\quad \valpha\notin C,\G}
{C\et P;\G\th t\au T}$\\[2mm]

($\au$$\ex$elim)\quad
$\cfrac{C;\G\th t\au \ex\valpha PT
\quad D;\G,x:T\th u\au U
\quad x\notin\G
\quad \valpha\notin C,\G}
{C\et \ex\valpha P\et \all\valpha(P\sgt D);
\G\th \open x=t\dans u\au \ex\valpha PU}$\\[5mm]


($\ad$abs)\quad
$\cfrac{C;\G,x:T\th u\ad U\quad x\notin\G}
{C;\G\th \l xu\ad T\A U}$\\[2mm]

($\ad$if)\quad
$\cfrac{C;\G\th t\ad\ex\alpha\bool^\alpha
\quad D;\G\th u\ad T\quad E;\G\th v\ad T
\quad T~\ex\mbox{-basic}}
{C\et D\et E;\G\th \si t\alors u\sinon v\ad T}$\\[2mm]

($\ad$$\all$intro)\quad
$\cfrac{C;\G\th t\ad T\quad \valpha\notin\G}
{\ex\valpha P\et \all\valpha(P\sgt C);\G\th t\ad \all\valpha PT}$\\[2mm]

($\ad$$\all$elim)\quad
$\cfrac{C;\G\th t\au \all\valpha PT}
{C\et P_\valpha^\va;\G\th t\ad T_\valpha^\va}$\\[2mm]

($\ad$$\ex$intro)\quad
$\cfrac{C;\G\th t\ad T \quad \valpha\notin\G}
{\ex\valpha(C\et P);\G\th t\ad \ex\valpha PT}$\\[2mm]

($\ad$$\ex$elim)\quad
$\cfrac{C;\G\th t\au \ex\valpha PT
\quad D;\G,x:T\th u\ad U
\quad \valpha\notin C,\G,U}
{C\et \ex\valpha P\et \all\valpha(P\sgt D);
\G\th \open x=t\dans u\ad U}$\\[2mm]

($\ad$sub)\quad
$\cfrac{C;\G\th t\au T'}
{C\et \J{T'\le T};\G\th t\ad T}$\\[2mm]
\vsp\end{minipage}}
\end{center}
\end{figure}


\begin{theorem}
Consider the rules of Figure \ref{fig-infer}. If $C;\G\th^? t:T$, then
$C$ is satisfiable and $C;\G\th t:T$.
\end{theorem}

\begin{proof}
First, one can easily check that, for every rule, if the constraint in
the conclusion is satisfiable, then the constraints in the premises
are satisfiable too. Then, we prove that, if $C$ is satisfiable and
$C;\G\th t\au T$ or $C;\G\th t\ad T$, then $C;\G\th t:T$. We only
detail some cases.

\begin{lst}{}







\item[($\au$$\ex$elim)]
Let $E=C\et\ex\valpha P\et\all\valpha(P\sgt D)$. Since $E\sgt C$ and
$(E\et P)\sgt D$, by induction hypothesis and weakening, $E;\G\th
t:\ex\valpha PT$ and $E\et P;\G\th u:U$. Since $(E\et P)\sgt P$, by
($\ex$intro), $E\et P;\G\th u:\ex\valpha PU$. Since $E\sgt\ex\valpha P$
and $\valpha\notin\ex\valpha PU$, by ($\ex$elim), $E;\G\th \open
x=t\dans u:\ex\valpha PU$.



\item[($\ad$$\all$intro)]
Let $E=\ex\valpha P\et\all\valpha(P\sgt C)$. Since $(E\et P)\sgt C$, by
induction hypothesis and weakening, $E\et P;\G\th t:T$. Since
$E\sgt\ex\valpha P$, we can conclude by ($\all$intro).

\item[($\ad$$\all$elim)]
Let $E=C\et P_\valpha^\va$. By induction hypothesis and weakening,
$E;\G\th t:\all\valpha PT$. Since $E\sgt P_\valpha^\va$, we can
conclude by ($\all$elim).

\item[($\ad$$\ex$intro)]
Let $E=\ex\valpha(C\et P)$. Since $E$ is satisfiable, $C$ is
satisfiable too. By completeness, there is $a$ such that
$F=C_\valpha^\va\et P_\valpha^\va$ is satisfiable. By induction
hypothesis, $C;\G\th t:T$. By substitution and weakening, $F;\G\th
t:T_\valpha^\va$. Since $F\sgt P_\valpha^\va$, by ($\ex$intro),
$F;\G\th t:\ex\valpha PT$. Since $E\sgt F$, we can conclude by
weakening.

\item[($\ad$$\ex$elim)]
Let $E=C\et\ex\valpha P\et\all\valpha(P\sgt D)$. Since $E\sgt C$ and
$(E\et P)\sgt D$, by induction hypothesis and weakening, $E;\G\th
t:\ex\valpha PT$ and $E\et P;\G\th u:U$. Since $E\sgt\ex\valpha P$ and
$\valpha\notin U$, by ($\ex$elim), $E;\G\th \open x=t\dans u:U$.\qed

\end{lst}
\end{proof}


\section{Termination proof technique}
\label{sec-sn}

In this section, we present a general method for proving the strong
normalization of $\b$-reduction and rewriting on well-typed terms. It
is based on Tait's method for proving the strong normalization of
$\b$-reduction \cite{tait72lc}. The idea is to interpret types by
particular sets of strongly normalizing terms, called saturated, and
prove that every well-typed term belongs to the interpretation of its
type.


Following \cite{abel04ita}, we define the {\em
weak-head-$\b$-reduction relation} $\abwh$ as the relation such that
$E[t]\abwh E[u]$ iff $t\abh u$ and $E\in\cE$, where the set of {\em
elimination contexts} $\cE$ is inductively defined as follows:
\[
E\in\cE ::= [] ~|~ E \, t ~|~ \fst \, E ~|~ \snd \, E
\]


\begin{definition}[Saturated sets]
The set $\SAT$ of {\em saturated sets} is the set of all the sets of
terms $S$ such that:
\begin{enumi}{}
\item If $t\in S$, then $t\in\SN$.
\item If $t\in S$ and $t\a t'$, then $t'\in S$.
\item If $E[x]\in\SN$, then $E[x]\in S$.
\item If $t\in\SN$, $t\abh t'$ and $E[t']\in S$, then $E[t]\in S$.
\end{enumi}
We also define the following operations on sets of terms:
\begin{lst}{--}
\item $S_1\A S_2=\{t\in\cT~|~\all u\in S_1,\, tu\in S_2\}$
\item $S_1\times S_2=\{t\in\cT~|~\fst t\in S_1\et \snd t\in S_2\}$
\end{lst}
Let $\cN$ be the set of terms of the form $\ff\vt$, $\si t\alors
u\sinon v$, $\fst t$ or $\snd t$. A saturated set $S$ has the {\em
neutral term property} if $s\in S$ whenever $s\in\cN$ and $\red{s}\sle
S$.
\end{definition}


\begin{lemma}
$\SAT$ is a complete lattice for inclusion with $\bigcup$ as lub,
$\bigcap$ as glb and $\SN$ as greatest element. It is also stable by
$\A$ and $\times$.
\end{lemma}

\comment{
\begin{proof}
One can easily check that $|SAT$ is closed by union and intersection,
and $\SN\in\SAT$. We now prove that $S_1\A S_2\in\SAT$ whenever
$S_1,S_2\in\SAT$.
\begin{enumi}
\item 
\item 
\item 
\item
\end{enumi}

We now prove that $S_1\times S_2\in\SAT$ whenever $S_1,S_2\in\SAT$.
\begin{enumi}
\item 
\item 
\item 
\item\qed
\end{enumi}
\end{proof}
}

All this is more or less well known. See for instance
\cite{abel04ita}. The key difference with the first author work
\cite{blanqui04rta} is that we use saturated sets instead of
reducibility candidates. See \cite{gallier90book} for a comparison
between the two kinds of sets. With reducibility candidates, (4) is
replaced by the neutral term property.

Reducibility candidates are saturated but the converse does not hold
since candidates are not stable by union. Hence, with candidates,
$\ex\valpha PT$ cannot be interpreted as an union, which is essential
if one wants to interpret $\fB^\ka$ as the set of terms of size $\ka$
in order to give precise types to function symbols.

However, reducibility candidates extend well to rewriting and
polymorphism since, for proving that $\ff\vt\in S$, it suffices to
prove that $\red{\ff\vt}\sle S$. In Lemma \ref{lem-neutr}, we prove
that this property still holds with saturated sets when $S$ is the
interpretation of an existentially quantified basic type.


\begin{definition}[Interpretation of types]
A base type interpretation is a function $I$ which, to every pair
$(\fB,\ka)$ with $\fB\neq\bool$, associates a set
$I_{\fB}^\ka\in\SAT$. We extend $I$ to $\bool$ by taking
$I_\bool^\ka=\{t\in\SN~|~\nf{t}\neq\ka^*\}$. Given such an
interpretation, types are interpreted by saturated sets as follows:

\begin{lst}{--}
\item $\I{\fB^a}^I_\mu= I_\fB^{a\mu}$
\item $\I{U\times V}^I_\mu= \I{U}^I_\mu\times \I{V}^I_\mu$
\item $\I{U\A V}^I_\mu= \I{U}^I_\mu\A \I{V}^I_\mu$
\item $\I{\all\valpha PT}^I_\mu=
\bigcap_{\mu+_\valpha^\vka\models P} \I{T}^I_{\mu+_\valpha^\vka}$
if~ $\th\!\ex\valpha P$, $\I{\all\valpha PT}^I_\mu=\SN$ otherwise
\item $\I{\ex\valpha PT}^I_\mu=
\bigcup_{\mu+_\valpha^\vka\models P} \I{T}^I_{\mu+_\valpha^\vka}$
if~ $\th\!\ex\valpha P$, $\I{\ex\valpha PT}^I_\mu=\bigcap\SAT$ otherwise
\end{lst}

\noindent
Let $I_\fB^\w=\I{\ex\alpha\fB^\alpha}$. A symbol $\fs\in\cC\cup\cF$ is
{\em computable} if $\fs\in\I{\tau_\fs}^I$. A pair $(\mu,\s)$ is {\em
valid} for $C;\G$, written $(\mu,\s)\models C;\G$, if $\mu\models C$
and, for all $x\in\dom(\G)$, $x\s\in\I{x\G}^I_\mu$. A base type
interpretation $I$ is valid if every constructor is computable and,
for every $\ex$-basic type $T$, $\I{T}^I_\mu$ has the neutral term
property.
\end{definition}

Note that $I_\bool^\ka\in\SAT$ has the neutral term property and
$\I{T\vphi}^I_\mu=\I{T}^I_{\vphi\mu}$.


\begin{theorem}
\label{thm-comp}
Assume that $I$ is a valid base type interpretation and every
$\ff\in\cF$ is computable. If $C;\G\th t:T$ and $(\mu,\s)\models
C;\G$, then $t\s\in\I{T}^I_\mu$.
\end{theorem}

\begin{proof}
By induction on $C;\G\th t:T$. We only detail some cases.

\begin{lst}{}


\item[(abs)]
We must prove that $s=(\l xu)\s\in\I{T\A U}^I_\mu$. Wlog, we can
assume that $x\notin\s$. Then, $s=\l x(u\s)$. Let
$t\in\I{T}^I_\mu$. We must prove that $st\in\I{U}^I_\mu$. By induction
hypothesis, $u\s\in\I{U}^I_\mu$. Let now $\s'=\s+_x^t$. Since
$(\mu,\s')\models C;\G,x:T$, by induction hypothesis,
$u\s'\in\I{U}^I_\mu$. Hence, $st\in\SN$ since, by induction on
$(u\s,t)$ with $\a\lex$ as well-founded ordering,
$\red{st}\sle\SN$. Therefore, $st\in\I{U}^I_\mu$ since $st\abh
u\s'\in\I{U}^I_\mu$ and $st\in\SN$.



\item[(if)]
Let $s=(\si t\alors u\sinon v)\s$. By induction hypothesis, $t\s\in
I_\bool^\w$ and $t_i\s\in\I{T}^I_\mu$. Since $s\in\cN$ and $T$ is an
$\ex$-basic type, by the neutral term property, it suffices to prove
that $\red{s}\sle\I{T}^I_\mu$. This follows by induction on
$(t\s,u\s,v\s)$ with $\a\lex$ as well-founded ordering.





\item[($\ex$elim)]
We must prove that $s=(\open x=t\dans u)\s\in\I{U}^I_\mu$. Wlog, we
can assume that $x\notin\s$. Then, $s=\open x=t\s\dans u\s$. Let
$\s'=\s_x^{t\s}$. By induction hypothesis, $t\s\in\I{\ex\valpha
PT}^I_\mu$. Since $\th C\sgt\ex\valpha P$, there is $\vka$ such that
$\mu+_\valpha^\vka\models P$ and
$t\s\in\I{T}^I_{\mu+_\valpha^\vka}$. Therefore, by induction
hypothesis, $u\s'\in\I{U}^I_{\mu+_\valpha^\va}=\I{U}^I_\mu$.

\item[(sub)] By induction on $T$ and $T'$, one can easily prove
that $\I{T}^I_\mu\sle\I{U}^I_\mu$ whenever $\mu\models\J{T\le U}$.\qed
\end{lst}
\end{proof}


\begin{corollary}
Assume that $I$ is a valid base type interpretation and every
$\ff\in\cF$ is computable. Then, $\a$ is strongly normalizing on
$\L(\tau)$.
\end{corollary}



\begin{corollary}
Assume that, for all $\fs\in\cC\cup\cF$, $\tau_\fs$ is of the form
$\vT\A\all\valpha\vec\fB^\valpha\A T$ with $\vT$ simple, $B$ basic and
$T$ an $\ex$-basic type. If every symbol is computable, then $\a$ is
strongly normalizing on $\o\L(\o\tau)$.
\end{corollary}

\begin{proof}
It suffices to prove that, for all $\fs$,
$\fs\in\I{\o{\tau_\fs}}^I$. We have
$\o{\tau_\fs}=\vT\A\vec\fB\A\o{B}$. Let $\vt\in\I\vT^I$ and $\vu\in
I_{\vec\fB}^\w$. We must prove that $\ff\vt\vu\in\I{\o{B}}^I$. There
is $\valpha\mu$ such that $\vu\in I_{\vec\fB}^{\valpha\mu}$. Assume
that $T=\all\vdelta\vP B$. Since
$\ff:\vT\A\all\valpha\vec\fB^\valpha\A T$ is computable,
$\ff\vt\vu\in\I{T}^I_\mu=
\bigcup_{\mu+_\vdelta^\kd\models\vP}\I{B}^I_{\mu+_\vdelta^\kd}$. Let
$\nu=\mu+_\vdelta^\kd\models\vP$. We are left to prove that
$\I{B}^I_\nu\sle\I{\o{B}}^I$. We proceed by induction on $B$.\qed

\end{proof}


\section{Termination criterion}
\label{sec-crit}

We now provide conditions to obtain the computability of defined symbols.

A {\em precedence} is a quasi-ordering $\ge$ whose strict part
${>}={\ge\moins\le}$ is well-founded. Let ${\simeq}={\ge\cap\le}$ be
its associated equivalence relation. We assume given a precedence
$\ge_\cB$ on $\cB$ and a precedence $\ge_\cF$ on $\cF$. We are going
to define some base type interpretation and prove that every function
symbol is computable by induction on these precedences.\\


{\bf Assumption:} For all $\fc\in\cC$, we assume that $\tau_\fc$ is of
the form\footnote{The order of types is not relevant.  We take this
order for the sake of simplicity.}
$\vec\fC\A\all\valpha\vec\fB^\valpha\A\fB^a$ with $\vec\fC<_\cB\fB$,
$\vec\fB\simeq_\cB\fB$, $a=0$ if $|\valpha|=0$, and $a=1+max(\valpha)$
if $|\valpha|>0$.

\begin{example}
The type $\nat$ of natural numbers has constructors $\ms{0}:\nat^0$
and $\fs:\all\alpha\nat^\alpha\A\nat^{\alpha+1}$. The type $\liste$ of
lists has constructors $\nil:\liste^0$ and
$\cons:\nat\A\all\alpha\liste^\alpha\A\liste^{\alpha+1}$. The type
$\tree$ of binary trees has constructors $\leaf:\nat\A\tree^0$ and
$\node:\all\alpha\b\tree^\alpha\A\tree^\b\A\tree^{1+max(\alpha,\b)}$.
\end{example}


We define the base type interpretation as follows:

\begin{lst}{--}
\item $I_\fB^0=\{t\in\SN~|~\all\fc:\vec\fC\A\all\valpha\vec\fB^\valpha\A\fB^a,\,
\all\vt\vu,\,|\vt|=|\vec\fC|\et |\vu|=|\valpha|\et\\
t\a^*\fc\vt\vu\A \vt\in I_{\vec\fC}^\w\et |\valpha|=a=0\}$
\item $I_\fB^{\ka+1}=\{t\in\SN~|~\all\fc:\vec\fC\A\all\valpha\vec\fB^\valpha\A\fB^a,\,
\all\vt\vu,\,|\vt|=|\vec\fC|\et |\vu|=|\valpha|\et\\
t\a^*\fc\vt\vu\A \vt\in I_{\vec\fC}^\w\et a=1+max(\valpha)\et
(\ex\vec\kb)\,\ka=max(\vec\kb)\et \vu\in I_{\vec\fB}^{\vec\kb}\}$
\end{lst}


\begin{lemma}
\label{lem-neutr}
$I$ is a valid base type interpretation.
\end{lemma}

\begin{proof}
One can easily check that $I_\fb^\ka$ is saturated and that every
constructor is computable. We now prove that $\I{T}^I_\mu$ has the
neutral term property whenever $T$ is $\ex$-basic.

We first remark that, if $t\in\SN$ and $t\a^* t'\in I_\fB^\ka$, then
$t\in I_\fB^\ka$. We prove it by induction on $(\fB,\ka)$ with
$(>_\cB,>_{\cD_{\kap_\cB}})\lex$ as well-founded ordering. Let
$\fc:\vec\fC\A\all\valpha\vec\fB^\valpha\A\fB^a$, $\vt$ and $\vu$ such
that $|\vt|=|\vec\fC|$, $|\vu|=|\valpha|$ and $t\a^*\fc\vt\vu$. By
confluence, $t'\a^*\fc\vt'\vu'$ with $\vt\vu\a^*\vt'\vu'$. We proceed
by case on $\ka$.
\begin{lst}{--}
\item $\ka=\dtrue$. Then, $t'\not\a^*\false$. Hence, $t\not\a^*\false$
and $t\in I_\fB^\ka$.
\item $\ka=\dfalse$. Idem.
\item $\ka=0$. Since $t'\in I_\fB^\ka$, $\vt'\in I_{\vec\fC}^\w$
and $|\valpha|=a=0$. Since $\vec\fC<_\cB\fB$, by induction hypothesis,
$\vt\in I_{\vec\fC}^\w$. Thus, $t\in I_\fB^\ka$.
\item $\ka>0$. Since $t'\in I_\fB^\ka$, $\vt'\in I_{\vec\fC}^\w$,
$a=1+max(\valpha)$ and there are $\vec\kb$ such that
$\ka=1+max(\vec\kb)$ and $\vu'\in I_{\vec\fB}^{\vec\kb}$. Since
$\vec\fC<_\cB\fB$ and $\vec\kb<\ka$, by induction hypothesis, $\vt\in
I_{\vec\fC}^\w$ and $\vu\in I_{\vec\fB}^{\vec\kb}$. Thus, $t\in
I_\fB^\ka$.
\end{lst}

\noindent
Let now $T=\ex\valpha\vP B$ be an $\ex$-basic type. We have
$S=\bigcup_{\mu+_\valpha^\vka\models\vP}\I{B}^I_{\mu+_\valpha^\vka}$. We
first prove that there are $\vka$ such that
$\nu=\mu+_\valpha^\vka\models\vP$ and $\red{s}\sle S'=\I{B}^I_\nu$. If
$\red{s}=\vide$, this is immediate. So, assume that there is
$t\in\red{s}$. Since $t\in S$, there are $\vka$ such that
$\nu=\mu+_\valpha^\vka\models\vP$ and $t\in S'=\I{B}^I_\nu$. Let now
$u\in\red{s}$. By confluence, there is $v$ such that $t,u\a^*
v$. Since $t\in S'$, we have $v\in S'$. Thus, $u\in S'$ too. Hence,
$\red{s}\sle S'$.

We now prove that $s\in S'$ whenever $\red{s}\sle S'$ by induction on
$B$.\qed

\end{proof}


\begin{figure}[ht]
\centering\caption{Matching constraints\label{fig-pat}}\vsp
\fbox{\begin{minipage}{12cm}\centering\vsp
(1)\quad $\alpha=\vep_x;x:\fB^{\vep_x}\leadsto x:\fB^\alpha$\\[2mm]

(2)\quad $\cfrac{\fc:\vT\A\fB^0\quad \fB\neq\bool}
{\alpha=0;\vx:\vT\leadsto \fc\vx:\fB^\alpha}$
\quad
(2')\quad $\cfrac{\fc:\bool^{\fc^*}}
{\alpha=\fc^*;\vide\leadsto \fc:\bool^\alpha}$\\[2mm]

(3)\quad $\cfrac{\begin{array}{c}
c:\vT\A\all\valpha\vec\fB^\valpha\A\fB^{1+max(\valpha)}\quad
\valpha=\va;\vec\G\leadsto \vu:\vec\fB^\valpha\quad
\alpha\notin\valpha\\
\vx:\vT,\vec\G \mbox{ are compatible}\\
\end{array}}
{\alpha=1+max(\va);\vx:\vT,\vec\G\leadsto \fc\vx\vu:\fB^\alpha}$
\vsp\end{minipage}}
\end{figure}


\begin{lemma}
\label{lem-acc}
We assume given an injection $\vep$ from term variables to size
variables. Consider the rules of Figure \ref{fig-pat}. If
$\alpha=a;\G\leadsto t:\fB^\alpha$ and $t\s\in I_\fB^{\alpha\mu}$,
then there is $\nu$ such that $(\mu+\nu,\s)\models\alpha=a;\G$.
\end{lemma}

\begin{proof}
We say that $\ka$ is minimal for $t\in\I\fB^\w$ if $t\in\I\fB^\ka$
and, for all $\kb<\ka$, $t\notin\I\fB^\kb$. We prove the lemma by
induction on $\alpha=a;\G\leadsto t:\fB^\alpha$ with the additional
requirement that $\nu$ is minimal whenever $\mu$ so is.

\begin{lst}{}
\item[(1)] It suffices to take $\vep_x\nu=\alpha\mu$.

\item[(2) and (2')] It suffices to take $\nu=\vide$.

\item[(3)] We have $t\s=c\vx\s\vu\s$. Thus, $\mu$ is minimal,
$x\s\in\I\vT$ and there is $\mu'$ minimal such that $\vu\s\in
I_{\vec\fB}^{\valpha\mu'}$ and $\alpha\mu=1+max(\valpha\mu')$. Now, by
induction hypothesis, there are $\vec\nu$ minimal such that
$(\mu'+\vec\nu,\s)\models\valpha=\va;\vec\G$. Since $\vec\nu$ are
minimal, if $x\s\in I_{\fB_i}^{\vep_x\nu_i}\cap
I_{\fB_j}^{\vep_x\nu_j}$, then $\vep_x\nu_i=\vep_x\nu_j$. Thus, we can
define $\nu=\S\vec\nu$. Since $\nu$ is minimal, we are left to prove
that $(\mu+\nu,\s)\models\alpha=1+max(\va);\vec\G$. First, we have
$\mu+\nu\models\alpha=1+max(\va)$ since
$\alpha\mu=1+max(\valpha\mu')=1+max(\va\nu)$. Second, let $x\in
u_i$. Then, $x\s\in\I{x\G}^I_{\nu_i}=\I{x\G}^I_{x\nu}$.\qed
\end{lst}
\end{proof}


\begin{theorem}[Termination criterion]
\label{thm-crit}
Assume that, for every $\ff\in\cF$:
\begin{enumi}{}
\item\label{item-type}
$\tau_\ff$ is of the form $\vT\A\all\valpha\vec\fB^\valpha\A T$ with
$T$ an $\ex$-basic type;
\item\label{item-measure}
there is a constraint $(\vbeta<_\ff\valpha)$ such that the ordering
$\succ_\ff$ defined by $\valpha\mu\succ_\ff\vbeta\mu$ iff
$\mu\models\vbeta<_\ff\valpha$ is well-founded;
\item for every $\fg\simeq_\cF\ff$, $\tau_\fg$ is of the form
$\vU\A\all\valpha\vec\fB^\valpha\A U$ and $<_\ff=<_\fg$;
\end{enumi}

\noindent
and, for every rule $\vt=\vec\fc\sgt l\a r$ defining $\ff$:
\begin{enumi}{}\setcounter{enumi}{3}
\item $l$ is of the form $\ff\vx\vl$ with $|\vx|=|\vT|$ and $|\vl|=|\valpha|$;
\item\label{item-matching}
there are $\vec\G$ compatible and $\va$ such that
$\valpha=\va;\vec\G\leadsto\vl:\vec\fB^\valpha$;
\item every symbol occurring in $r$ is $\le_\cF f$;
\item\label{item-cond}
$\valpha=\va;\vx:\vT,\vec\G\th_{\tau^<} \vt:\bool^\vb$;
\item\label{item-rhs}
$\vb=\vec\fc^*;\valpha=\va;\vx:\vT,\vec\G\th_{\tau^<} r:T$.
\end{enumi}

\noindent
where:
\begin{enumi}{}\setcounter{enumi}{8}
\item for every $\fg<_\cF\ff$, $\tau^<_\fg=\tau_\fg$;
\item\label{item-termin}
for every $\fg\simeq_\cF\ff$,
$\tau^<_\fg=\vU\A\all\valpha'(\valpha'<_\ff\valpha)\vec\fB^{\valpha'}\A
U$ with $\valpha'\notin\valpha$ whenever
$\tau_\fg=\vU\A\all\valpha'\vec\fB^{\valpha'}\A U$.
\end{enumi}

\noindent
Then, $\a$ is strongly normalizing on $\L(\tau)$ and $\o\L(\o\tau)$.
\end{theorem}

\begin{proof}
We must prove that, for all
$\ff:\vT\A\all\valpha\vec\fB^\valpha\A T$, $\vt\in\I\vT$,
$\mu$ and $\vu\in I_{\vec\fB}^{\valpha\mu}$,
$\ff\vt\vu\in\I{T}^I_\mu$. We proceed by induction on
$(\ff,\alpha\mu,\vt\vu)$ with $(>_\cF,\succ_\ff,\a\lex)\lex$ as
well-founded ordering. By Lemma \ref{lem-neutr}, it suffices to prove
that $\red{s}\sle S$. If the reduction takes place in $\vt\vu$, we
conclude by induction hypothesis. Assume now that there are
$\ff\vx\vl\a r\in\cR$ and $\s$ such that $\vx\s=\vt$ and
$\vl\s=\vu$. We must prove that $r\s\in\I{T}^I_\mu$. After Lemma
\ref{lem-acc}, since $\vec\G$ are compatible, there is $\nu$ such that
$(\mu+\nu,\s)\models\valpha=\va;\vec\G$. By induction hypothesis, for
all $g\le_\cF f$, $g\in\I{\tau^<_\fg}$ (considering $\valpha$ as
constants interpreted by $\valpha\mu$). Thus, letting $\eta=\mu+\nu$,
by Theorem \ref{thm-comp}, we have $\vt\s\in I_\bool^{\vb\eta}$. Since
$t\s\a^*\vec\fc\in I_\bool^{\vec\fc^{**}}$, we have
$\vb\eta=\vec\fc^{**}$. Thus, $\eta\models\vb=\vec\fc^*$ and, by
Theorem \ref{thm-comp} again, $r\s\in\I{T}^I_\eta=\I{T}^I_\mu$.\qed
\end{proof}

The size variables $\valpha$ in the type of $\ff$ (\ref{item-type})
represents the sizes of the recursive arguments of $\ff$. The
user-defined predicate $<_\ff$ in (\ref{item-measure}) expresses the
measure that must decrease in recursive calls. One can for instance
take lexicographic or multiset comparisons together with linear
combinations of the arguments. The condition (\ref{item-matching})
provides the constraints on $\valpha$ when a term matches the rule
left hand-side $l=\ff\vx\vl$. The condition (\ref{item-cond}) implies
that the terms $\vt$ are terminating whenever the arguments of the
left hand-side so are. The condition (\ref{item-rhs}) implies that the
right hand-side is terminating whenever the arguments of the left
hand-side so are and $\vt\a^*\vec\fc$. The fact that $\vt\a^*\vec\fc$
is expressed by the additional constraint $\vb=\vec\fc^*$.
Termination is ensured by doing type-checking in the system
$\th_{\tau^<}$ where, by condition (\ref{item-termin}), function
symbols equivalent to $\ff$ can only be applied to arguments smaller
than $\valpha$ in $<_\ff$. This is in contrast with
\cite{blanqui04rta} where a new type system (called the computability
closure) restricting the use of (app) must be introduced.


\begin{example}
\label{ex-pivot}
We detail the criterion with the second rule of $\pivot$ given in the
introduction. Let $r$ be the right-hand side of the rule and $u$
(resp. $v$) be the first (resp. second) branch of $\si\!$ in $r$.

We take $\pivot:\nat\A\all\alpha\liste^\alpha\A T(\alpha)$ with
$T(\alpha)=\ex\b\g(\alpha=\b+\g)\liste^\b\times\liste^\g$, ${<_\ff} =
{<}$, ${\succ_\ff}={>_\bN}$ and $\inf:\nat\A\nat\A\bool$. Let
$\G=y:\nat,l:\liste^\d$ and $\D=x:\nat,\G$.

Matching constraint: $\alpha=\d+1;\G\leadsto \cons y\,l:\liste^\alpha$
(we take $\vep_l=\d$).

We must check that ${\alpha=\d+1};\D\th r: T(\alpha)$ with
$\pivot:\nat\A\all\alpha'(\alpha'<\alpha)\liste^{\alpha'}\A
T(\alpha')$. Let $\D=\G,z:\liste^\b\times\liste^\g$.

One can easily check that $\d<\alpha;\G\th \pivot x\,l\au T(\delta)$,
$\top;\D\th \inf y\,x\au\bool$, $\top;\D\th u\au
\liste^{\b+1}\times\liste^\g$, $\top;\D\th v\au
\liste^\b\times\liste^{\g+1}$.

Thus, by ($\ad$sub), $\b+1=\b'\et\g=\g';\D\th u\ad
\liste^{\b'}\times\liste^{\g'}$ and $\b=\b'\et\g+1=\g';\D\th u\ad
\liste^{\b'}\times\liste^{\g'}$.

By ($\ad$$\ex$intro), $D;\D\th u\ad T(\alpha)$ where
$D=\ex\b'\g'(\b+1=\b'\et\g=\g'\et\alpha=\b'+\g')$, and $E;\D\th v\ad
T(\alpha)$ where $E=\ex\b'\g'(\b=\b'\et\g+1=\g'\et\alpha=\b'+\g')$.
Note that $D\equiv E\equiv (\alpha=\b+\g+1)$.

By ($\ad$if), $\alpha=\b+\g+1;\D\th \si(\inf y\,x)\alors u\sinon v:T(\alpha)$.

By ($\ad$$\ex$elim), $F;\G\th r\ad T(\alpha)$ where
$F=\delta<\alpha\et (\ex\b\g(\alpha=\b+\g))\et
(\all\b\g(\delta=\b+\g\sgt\alpha=\b+\g+1))$.

Therefore, ${\alpha=\d+1};\D\th r: T(\alpha)$ if $\th\alpha=\d+1\sgt
F$, which is true.
\end{example}


\begin{example}
Consider the following definition of Mc Carthy's 91 function:

\begin{center}
$\begin{array}{r@{~=~}l@{~\sgt~}r@{~\a~}l}
\inf x\,100 & \true & \ff\,x & \ff\,(\ff\,(\plus x\,11))\\
\inf x\,100 & \false & \ff\,x & \minus x\,10\\
\end{array}$
\end{center}

\noindent
We assume that $\cA$ contains $le:nat\times nat\A bool$ interpreted as
expected.

We assume that
$\inf:\all\alpha\b\nat^\alpha\A\nat^\b\A\bool^{le(\alpha,\b)}$,
$\plus:\all\alpha\b\nat^\alpha\A\nat^\b\A\nat^{\alpha+\b}$,
$\minus:\all\alpha\b\nat^\alpha\A\nat^\b\A\ex\g P\nat^\g$ with
$P=({\alpha\le\b}\et{\g=0})\ou({\alpha>\b}\et{\alpha=\b+\g})$, and
$\ff:\all\alpha\nat^\alpha\A\ex\b Q\nat^\b$ with $Q=({\alpha\le
100}\et{\b=91})\ou({\alpha>100}\et{\alpha=\b+10})$. Taking
$\G=x:\nat^\alpha$, we get that $\top;\G\th \inf
x\,100:\bool^{le(\alpha,100)}$. The condition $le(\alpha,100)=\atrue$
is equivalent to $\alpha\le 100$, hence the termination.
\end{example}


\section{Conclusion and future work}

We extended the simply typed part of \cite{blanqui04rta} with
conditional rewriting and explicit quantifications and constraints
over size annotations. This allows to precisely describe the relation
between the size of the output of a function and the size of its
inputs. This also provides a powerful termination criterion for the
combination of $\b$-reduction and higher-order conditional rewriting,
based on type-checking and constraint solving. To our knowledge, this
is the first termination criterion for higher-order conditional
rewriting taking into account conditions in termination. We plan to
extend this work in various directions:

\begin{lst}{--}
\item As in \cite{xi02hosc}, we did not consider constructors with
recursive arguments of higher-order type since this is already studied
in \cite{blanqui04rta}. The integration of both works should not
create too much difficulties. We already have preliminary results in
this direction.

\item The complexity of Presburger arithmetic is high.
Although it is not so important in our case since the constraints we
consider are small (rule right-hand sides are generally not very big
terms), it would be interesting to study the complexity in more
details, depending on the allowed size annotations.

\item Our long term goal is to extend the present work to
polymorphic and dependent type systems that serve as basis for
proof assistants like Coq, \eg the Calculus of Algebraic Constructions
\cite{blanqui05mscs}.

\item We assume that constrained types of function symbols
are given and check that they imply termination. It would be very
interesting to infer these constraints automatically.
\end{lst}


\end{document}